\documentclass[11pt]{article}

\usepackage[dvipdfm,
		colorlinks=true,
		linkcolor=red,
		urlcolor=black,
		bookmarks=true,
		bookmarksnumbered=true,
		bookmarkstype=toc]{hyperref}
\usepackage{amsmath}
\usepackage{amssymb}
\usepackage{amsbsy}
\usepackage{mathrsfs}
\usepackage{graphicx}
\usepackage{color}
\newcommand{\1}{\mbox{1}\hspace{-0.25em}\mbox{l}}

\title{%
	\vskip -3em
	\begin{flushright}{\small KOBE-TH-08-10}\end{flushright}
	\vskip 5em
	\textbf{\large Hierarchy of QM SUSYs on a Bounded Domain}
	\vskip 1em
}
\author{
Tomoaki Nagasawa$^{1,}$\thanks{\tt e-mail:\href{mailto:nagasawa@anan-nct.ac.jp}{nagasawa@anan-nct.ac.jp}},~~
Satoshi Ohya$^{2,}$\thanks{\tt e-mail:\href{mailto:ohya@kobe-u.ac.jp}{ohya@kobe-u.ac.jp}},~~
Kazuki Sakamoto$^{2,}$\thanks{\tt e-mail:\href{mailto:049d841n@stu.kobe-u.ac.jp}{049d841n@stu.kobe-u.ac.jp}},\\[1ex]
Makoto Sakamoto$^{3,}$\thanks{\tt e-mail:\href{mailto:dragon@kobe-u.ac.jp}{dragon@kobe-u.ac.jp}}~~~and~ 
Kosuke Sekiya$^{2,}$\thanks{\tt e-mail:\href{mailto:065s115n@stu.kobe-u.ac.jp}{065s115n@stu.kobe-u.ac.jp}}\\
\vspace{1ex}\\
\textit{\small $^{1}$Anan National College of Technology, 265 Aoki, Minobayashi, Anan 774-0017, Japan}\\
\textit{\small $^{2}$Graduate School of Science, Kobe University, 1-1 Rokkodai, Nada, Kobe 657-8501, Japan}\\
\textit{\small $^{3}$Department of Physics, Kobe University, 1-1 Rokkodai, Nada, Kobe 657-8501, Japan}
}
\date{%
\small (Dated: \today)
}

\begin{document}
\maketitle
%%%%% ABSTRACT %%%%%
\begin{abstract}
We systematically formulate a hierarchy of isospectral Hamiltonians in one-dimensional supersymmetric quantum mechanics on an interval and on a circle, in which two successive Hamiltonians form $\mathscr{N} = 2$ supersymmetry.
We find that boundary conditions compatible with supersymmetry are severely restricted.
In the case of an interval, a hierarchy of, at most, three isospectral Hamiltonians is possible with unique boundary conditions, while in the case of a circle an infinite tower of isospectral Hamiltonians can be constructed with two-parameter family of boundary conditions.
\end{abstract}
\newpage

%%%%% INTRODUCTION %%%%%
\section{Introduction} \label{sec:intro}
Although historically supersymmetric quantum mechanics (SUSY QM) was originally introduced by Witten \cite{Witten:1981} as a toy model of studying patterns of supersymmetry breakings, it was soon recognized that SUSY QM was interesting in its own right; for example, it provides a systematic description of categorizing analytically solvable potentials using the so-called shape invariance (see for review \cite{Cooper:1994}).
Schr\"odinger equations with shape invariant potentials can be solved algebraically with the aid of supersymmetry.
SUSY QM also appears in various contexts of physics; it is related to soliton physics \cite{Wang:1990,Grant:1993,Rodrigues:1998,GomesLima:2002,Dias:2002,deLima:2003} including inverse scattering problems \cite{Sukumar:1985,Sparenberg:1997,Baye:2004}, two-dimensional quantum field theories \cite{Feinberg:2003,Seeger:1998}, supersymmetric lattice models leaving time-direction continuous \cite{Bergner:2007}, integrable models such as the Calogero model and its application to black hole physics \cite{Calogero:1971,Sutherland:1972,Moser:1975,Gibbons:1998,Meljanac:2006}, and quantum mechanics with point singularities \cite{point_singularity1, point_singularity2}.

Recently it was shown that in higher dimensional gauge theories with extra compact dimensions there always exists an $\mathscr{N}=2$ quantum mechanical supersymmetry (QM SUSY) in the 4d spectrum; the Kaluza-Klein mass eigenvalue problems are equivalent to energy eigenvalue problems in $\mathscr{N} = 2$ SUSY QM \cite{Lim:2005}.
The $\mathscr{N}=2$ QM SUSY can be regarded as a remnant of the higher-dimensional gauge invariance, and plays an essential role to generate an infinite tower of massive spin-1 particles.
In Ref.\cite{hierarchy}, it was pointed out that a hierarchical mass spectrum can naturally arise in the context of a higher dimensional gauge theory with a warped metric and give a solution to the gauge hierarchy problem, in which the $\mathscr{N}=2$ QM SUSY turns out to play a crucial role.
Since the extra dimension is compactified, the corresponding supersymmetric quantum mechanical systems are of course constrained to bounded domains.
There, boundary conditions are very important not only for the infrared regime but also for the ultraviolet regime, and play an essential role to determine the 4d particle spectrum especially for the low energy levels or massless mode.
When the compactified dimension does not respect the translational invariance due to the presence of extended defects (branes or boundaries), boundary effects also play a significant role in the ultraviolet regime as boundary localized divergent terms \cite{Georgi:2000}.
Such localized ultraviolet divergences must be renormalized by field theory operators on the boundary and give rise to nontrivial renormalization group flows for brane localized theory \cite{Goldberger:2001,Milton:2001}.
Since any gauge invariant field theory possesses the $\mathscr{N}=2$ QM SUSY, the boundary conditions and the $\mathscr{N}=2$ QM SUSY must be compatible with each other.
In this paper we will address this issue from the supersymmetric quantum mechanics point of view:
we analyze the possible boundary conditions in one-dimensional $\mathscr{N}=2$ SUSY QM on a bounded domain $(0,L)$.

The analysis developed in \cite{Lim:2005} was extended to 5d gravity \cite{Lim:2007}.
In 5d gravity it was shown that \textit{two} $\mathscr{N}=2$ SUSYs are hidden in the 4d spectrum.
The two $\mathscr{N}=2$ SUSYs can be regarded as a remnant of higher-dimensional general coordinate invariance, and are needed in order for the ``Higgs'' mechanism to generate massive spin-2 particles; one of the two quantum mechanical SUSYs ensures the degeneracy between spin-2 and spin-1 excitations and the other between spin-1 and spin-0 excitations.
A crucial ingredient of this coexistence of two quantum mechanical SUSYs is the refactorization of Hamiltonians (Laplace operators).
In view of these facts it would be natural to guess that in a higher-dimensional spin-$N$ field theory there would exist $N$ $\mathscr{N}=2$ SUSYs in the 4d mass spectrum.
In this paper we will also investigate whether it is possible to construct such a hierarchy of $N$ SUSYs without conflicting with the boundary conditions.

The rest of this paper is organized as follows.
In Section \ref{sec:BC} we analyze the possible boundary conditions in $\mathscr{N}=2$ SUSY QM on a bounded domain $(0,L)$.
We show that the allowed boundary conditions in $\mathscr{N}=2$ SUSY QM is limited to the so-called scale-independent subfamily of the $U(2)$ family of boundary conditions \cite{Cheon:2000}.
In Section \ref{sec:refactorization} we construct a hierarchy of $N$ SUSYs by solving the refactorization condition.
The results coincide with the so-called isospectral deformations of the Hamiltonian \cite{Abraham:1980,Baye:1987,Amado:1988}.
In Section \ref{sec:hierarchy} we analyze the allowed boundary conditions of quantum mechanical system with $N$ SUSYs on an interval and on a circle separately and present a systematic prescription to construct a hierarchy
of isospectral Hamiltonians.
Section \ref{sec:conclusion} is devoted to conclusions and discussions.

%%%%% BOUNDARY CONDITIONS %%%%%
\section{Boundary conditions in $\mathscr{N}=2$ SUSY QM} \label{sec:BC}
Hermiticity of Hamiltonian is the basic principle in quantum theory; it leads to the unitarity of the S-matrix or the conservation of probability in the whole quantum system.
In one-dimensional non-supersymmetric quantum mechanics it is known that the most general boundary conditions consistent with the hermiticity of Hamiltonian are characterized by a $2\times2$ unitary matrix $U$ \cite{Cheon:2000}.
In one-dimensional $\mathscr{N}=2$ SUSY QM, however, supersymmetry imposes more severe constraints on the parameter space of this $U(2)$ family of boundary conditions.
As we will show below the possible boundary conditions consistent with $\mathscr{N}=2$ supersymmetry are limited to the so-called scale-independent subfamily of the $U(2)$ family of boundary conditions.

To begin with let us consider $\mathscr{N}=2$ SUSY QM on a finite domain $(0, L) \in \mathbb{R}$, whose Hamiltonians are given by\footnote
{%
$\mathscr{N} = 2$ supersymmetry will be transparent by introducing the following $2\times 2$ matrix operators
\begin{align}
\mathscr{H}
= 	\begin{bmatrix}
	H_{0} 	& 0 \\
	0 		& H_{1}
	\end{bmatrix}, \quad
(-1)^{F}
= 	\begin{bmatrix}
	1 	& 0 \\
	0 	& -1
	\end{bmatrix}, \quad
\mathscr{Q}_{1}
= 	\begin{bmatrix}
	0 		& Q_{0}^{\dagger} \\
	Q_{0} 	& 0
	\end{bmatrix}, \quad
\mathscr{Q}_{2}
= 	i(-1)^{F}\mathscr{Q}_{1}, \nonumber
\end{align}
which satisfy the standard $\mathscr{N} = 2$ supersymmetry algebra
\begin{align}
\{\mathscr{Q}_{i}, \mathscr{Q}_{j}\}
= 	2\delta_{ij}\mathscr{H}, \quad
[\mathscr{Q}_{i}, \mathscr{H}]
= 0, \quad
[(-1)^{F}, \mathscr{H}]
=0, \quad
\{(-1)^{F}, \mathscr{Q}_{i}\}
=0, \quad
i,j
= 1,2. \nonumber
\end{align}
}
%%%%%%%%%%%%%%%%%
\begin{subequations}
\begin{align}
H_{0}
&= 	Q_{0}^{\dagger}Q_{0}, \label{eq:2.2a}\\
H_{1}
&= 	Q_{0}Q_{0}^{\dagger}. \label{eq:2.2b}
\end{align}
\end{subequations}
The supercharge $Q_{0}$ and its adjoint $Q_{0}^{\dagger}$ are given by
\begin{subequations}
\begin{align}
Q_{0}
&= 	\frac{\mathrm{d}}{\mathrm{d}x} + W_{0}^{\prime}(x), \label{eq:2.3a}\\
Q_{0}^{\dagger}
&= 	- \frac{\mathrm{d}}{\mathrm{d}x} + W_{0}^{\prime}(x), \label{eq:2.3b}
\end{align}
\end{subequations}
where $W_{0}$ is a superpotential (or prepotential), which must be a real function in order to guarantee the hermiticity of the Hamiltonians, and prime ($\prime$) indicates the derivative with respect to $x$.
In terms of the zero-mode function $\phi_{0}^{(0)}$ satisfying the equation $Q_{0}\phi_{0}^{(0)} = 0$, the superpotential $W_{0}$ can be written as
\begin{align}
W_{0}(x) = -\ln\phi_{0}^{(0)}(x). \label{eq:2.4}
\end{align}
Supersymmetric relations are
\begin{subequations}
\begin{align}
Q_{0}\phi_{0}
&= 	\sqrt{E}\phi_{1}, \label{eq:2.5a}\\
Q_{0}^{\dagger}\phi_{1}
&= 	\sqrt{E}\phi_{0}, \label{eq:2.5b}
\end{align}
\end{subequations}
where $\phi_{0}$ and $\phi_{1}$ are eigenfunctions of $H_{0}$ and $H_{1}$, respectively, with the common energy $E$.
In this paper we will concentrate on a finite superpotential on the whole domain.
In other words, we require that $\phi_{0}^{(0)}$ has no zero point (or no node).

Next we will focus on the hermiticity of $H_{0}$ and then derive the allowed boundary conditions for $\phi_{0}$ and $\phi_{1}$ using the supersymmetric relations \eqref{eq:2.5a} \eqref{eq:2.5b}.
In physical language, the hermiticity of the Hamiltonian $H_{0}$ indicates the conservation of probability in the whole system $j_{0}(0) = j_{0}(L)$, where the probability current density $j_{0}$ is defined by $j_{0} = -i((\phi_{0}^{\ast})^{\prime}\phi_{0} - \phi_{0}^{\ast}\phi_{0}^{\prime})$.
It is more suitable for the following discussion to rewrite the probability current density into the following form
\begin{align}
j_{0}(x)
&= 	-i\bigl[(Q_{0}\phi_{0})^{\ast}(x)\phi_{0}(x) - \phi_{0}^{\ast}(x)(Q_{0}\phi_{0})(x)\bigr], \label{eq:2.6}
\end{align}
which follows from the real-valued superpotential.

There are two physically distinct cases:
\begin{enumerate}
	\item Case $j_{0}(0) = 0 = j_{0}(L)$.\\
		In this case the probability current density $j_{0}$ does not flow outside the domain and the probability is locally conserved.
		Hence the two ends of the domain $x = 0$ and $L$ are physically disconnected and we will refer to this case as an interval case.
	\item Case $j_{0}(0) = j_{0}(L) (\neq 0)$.\\
		In this case $j_{0}$ flows outside the domain but the probability is globally conserved as an entire system, which implies that the two ends of the domain are physically connected.
		Hence we will refer to this case as a circle case.
		Although in this case the end points $x = 0$ and $L$ are physically identified, there is no need the superpotential $W_{0}$ to be a periodic function; when the superpotential does not have a periodicity of $L$, there just arises some kind of singularity at the junction point $x=0$, which can be characterized by the boundary conditions just as in the point interactions \cite{Cheon:2000}.
\end{enumerate}
In the following subsections we will study these two cases separately.

%%%%% INTERVAL %%%%%
\subsection{Interval case: $j_{0}(0) = 0 = j_{0}(L)$} \label{subsec:interval}
We first investigate the condition $j_{0}(0) = 0 = j_{0}(L)$.
Note that the condition $j_{0}(x_{i}) = 0$ ($i = 1, 2$; $x_{1} = 0, x_{2} = L$) can be written as follows:
\begin{align}
\bigl|\phi_{0}(x_{i}) - iL_{0}(Q_{0}\phi_{0})(x_{i})\bigr|^{2}
&= 	\bigl|\phi_{0}(x_{i}) + iL_{0}(Q_{0}\phi_{0})(x_{i})\bigr|^{2}, \label{eq:2.1.1}
\end{align}
where $L_{0}$ is an arbitrary real constant of mass dimension $-1$, which is just introduced to adjust the mass dimension of the equation.
As we will see below $L_{0}$ is not a parameter characterizing the boundary conditions.

The above equation implies that the two complex numbers $\phi_{0}(x_{i}) - iL_{0}(Q_{0}\phi_{0})(x_{i})$ and $\phi_{0}(x_{i}) + iL_{0}(Q_{0}\phi_{0})(x_{i})$ are different from each other at most only in a phase factor.
Thus we can write
\begin{align}
\phi_{0}(x_{i}) - iL_{0}(Q_{0}\phi_{0})(x_{i})
&= 	{\rm e}^{i\theta_{i}}\bigl(\phi_{0}(x_{i}) + iL_{0}(Q_{0}\phi_{0})(x_{i})\bigr), \label{eq:2.1.2}
\end{align}
where $0\leq\theta_{i}<2\pi$, $i=1,2$.
When one considers a non-supersymmetric quantum mechanics, this is the end of the story by just replacing the supercharge $Q_{0}$ to the ordinary derivative $\mathrm{d}/\mathrm{d}x$, and the resulting boundary conditions are parameterized by the group $U(1)\times U(1)$, whose parameter space is a 2-torus $S^{1}\times S^{1}\simeq T^{2}$ \cite{Cheon:2000}.
However, supersymmetry severely restricts the allowed parameter space.
Using the supersymmetric relations \eqref{eq:2.5a} and \eqref{eq:2.5b} we find
\begin{subequations}
\begin{align}
\sin\left(\frac{\theta_{i}}{2}\right)\phi_{0}(x_{i})
+ L_{0}\cos\left(\frac{\theta_{i}}{2}\right)(Q_{0}\phi_{0})(x_{i})
&= 0, \label{eq:2.1.3a}\\
\sin\left(\frac{\theta_{i}}{2}\right)(Q_{0}^{\dagger}\phi_{1})(x_{i})
+ EL_{0}\cos\left(\frac{\theta_{i}}{2}\right)\phi_{1}(x_{i})
&= 0. \label{eq:2.1.3b}
\end{align}
\end{subequations}
Since the boundary conditions should not depend on the eigenvalue $E$ (otherwise the superposition of the quantum states becomes meaningless), 
the parameters $\theta_{i}$ ($i=1,2$) must be $0$ or $\pi$.
Thus in $\mathscr{N}=2$ SUSY QM on an interval the boundary conditions compatible with the supersymmetry are characterized by the discrete group $\mathbb{Z}_{2}\times \mathbb{Z}_{2}\subset U(1)\times U(1)$, which just consists of four 0-dimensional points $\{\mathrm{e}^{i0}, \mathrm{e}^{i\pi}\}\times\{\mathrm{e}^{i0}, \mathrm{e}^{i\pi}\} = \{1, -1\}\times \{1, -1\}$.
This result is consistent with the previous analyses of SUSY QM
with point singularities \cite{point_singularity1, point_singularity2}.
Now it is clear that the allowed boundary conditions can be categorized into the following $2\times2 = 4$ types:
\begin{subequations}
\begin{alignat}{2}
(\theta_{1}, \theta_{2}) &= (0, 0) & : &\quad
	\begin{cases}
	(Q_{0}\phi_{0})(0) = 0 = (Q_{0}\phi_{0})(L), \\
	\phi_{1}(0) = 0 = \phi_{1}(L);
	\end{cases} \\
(\theta_{1}, \theta_{2}) &= (\pi, \pi)~& :& \quad
	\begin{cases}
	\phi_{0}(0) = 0 = \phi_{0}(L), \\
	(Q_{0}^{\dagger}\phi_{1})(0) = 0 = (Q_{0}^{\dagger}\phi_{1})(L);
	\end{cases} \\
(\theta_{1}, \theta_{2}) &= (0, \pi) & : &\quad
	\begin{cases}
	(Q_{0}\phi_{0})(0) = 0 = \phi_{0}(L), \\
	\phi_{1}(0) = 0 = (Q_{0}^{\dagger}\phi_{1})(L);
	\end{cases} \\
(\theta_{1}, \theta_{2}) &= (\pi, 0) & : &\quad
	\begin{cases}
	\phi_{0}(0) = 0 = (Q_{0}\phi_{0})(L), \\
	(Q_{0}^{\dagger}\phi_{1})(0) = 0 = \phi_{1}(L).
	\end{cases}
\end{alignat}
\end{subequations}

%%%%% CIRCLE %%%%%
\subsection{Circle case: $j_{0}(0) = j_{0}(L) (\neq 0)$} \label{subsec:circle}
Next investigate the condition $j_{0}(0) = j_{0}(L) (\neq 0)$.
This condition can be written into the following form
\begin{align}
\left|
\Phi_{\phi_{0}}
-iL_{0}\sigma_{3}
\Phi_{Q_{0}\phi_{0}}
\right|^{2}
&=
\left|
\Phi_{\phi_{0}}
+iL_{0}\sigma_{3}
\Phi_{Q_{0}\phi_{0}}
\right|^{2}, \label{eq:2.2.1}
\end{align}
where for any function $f(x)$ the two-component boundary value vector $\Phi_{f}$ is defined as
\begin{align}
\Phi_{f}
:= 	\begin{bmatrix}
	f(0) \\
	f(L)
	\end{bmatrix}. \label{eq:2.2.2}
\end{align}
$\sigma_{3}$ is the third Pauli matrix: $\sigma_{3} = \mathrm{diag}(1, -1)$.
This equation shows that the squared length of the two-dimensional complex column vector
$\Phi_{\phi_{0}} -iL_{0}\sigma_{3}\Phi_{Q_{0}\phi_{0}}$
is equal to that of 
$\Phi_{\phi_{0}} +iL_{0}\sigma_{3}\Phi_{Q_{0}\phi_{0}}$,
which implies that these two vectors must be related by a two-dimensional unitary transformation.
Thus we can write
\begin{align}
\Phi_{\phi_{0}} -iL_{0}\sigma_{3}\Phi_{Q_{0}\phi_{0}}
&=
U\left(\Phi_{\phi_{0}} +iL_{0}\sigma_{3}\Phi_{Q_{0}\phi_{0}}\right), \label{eq:2.2.3}
\end{align}
where $U$ is an arbitrary $2\times2$ unitary matrix.
In one-dimensional non-supersymmetric quantum mechanics it is known that the most general boundary conditions are characterized by this $U(2)$ family \cite{Cheon:2000}.
In the following we shall determine the possible form of this unitary matrix compatible with supersymmetry and find the allowed subspace of the $U(2)$ family.

To this end we first apply the supersymmetric relations to the condition \eqref{eq:2.2.3}.
Using the supersymmetric relations \eqref{eq:2.5a} and \eqref{eq:2.5b} we find
\begin{subequations}
\begin{align}
(\1-U)\Phi_{\phi_{0}} -iL_{0}(\1+U)\sigma_{3}\Phi_{Q_{0}\phi_{0}}
&= \vec{0}, \label{eq:2.2.4a}\\
(\1-U)\Phi_{Q_{0}^{\dagger}\phi_{1}} -iEL_{0}(\1+U)\sigma_{3}\Phi_{\phi_{1}}
&= \vec{0}. \label{eq:2.2.4b}
\end{align}
\end{subequations}
Again since the boundary conditions should not depend of the eigenvalue $E$, the eigenvalues of the matrix $U$ must be $1$ or $-1$, which is equivalent to the condition $U^{2} = \1$.
Notice that any unitary matrix satisfying $U^{2} = \1$ can be spectrally decomposed using the projection operators $P_{+} = \frac{1}{2}(\1 + U)$ and $P_{-} = \frac{1}{2}(\1 - U)$, which satisfy $P_{+} + P_{-} = \1$, $(P_{\pm})^{2} = \1$ and $P_{\pm}P_{\mp} = 0$.
Multiplying these projection operators the above boundary conditions boil down to the following four independent conditions:
\begin{subequations}
\begin{align}
(\1-U)\Phi_{\phi_{0}}
&= 	\vec{0}, \label{eq:2.2.5a}\\
(\1+U)\sigma_{3}\Phi_{Q_{0}\phi_{0}}
&= 	\vec{0}, \label{eq:2.2.5b}\\
(\1-U)\Phi_{Q_{0}^{\dagger}\phi_{1}}
&= 	\vec{0}, \label{eq:2.2.5c}\\
(\1+U)\sigma_{3}\Phi_{\phi_{1}}
&= 	\vec{0}. \label{eq:2.2.5d}
\end{align}
\end{subequations}
Note that when $U=\1$ ($U=-\1$) these boundary conditions reduce to type ($0,0$) (type ($\pi, \pi$)) boundary conditions in the interval case and lead to $j_{0}(0) = 0 = j_{0}(L)$.
Thus in this circle case these two ``points'' $U = \1$ and $-\1$ have to be removed from the parameter space, from which we conclude that the two eigenvalues of $U$ must be $1$ and $-1$.
Such a unitary matrix can be written as follows:
\begin{align}
U = \vec{e}\cdot\vec{\sigma}, \label{eq:2.2.6}
\end{align}
where ${\vec \sigma}$ are the Pauli matrices and ${\vec e}$ is a unit vector, which can be parameterized as
\begin{align}
\vec{e} = (\cos\theta\sin\phi, \sin\theta\sin\phi, \cos\phi), \quad
0\leq\theta<2\pi, \quad
0\leq\phi\leq\pi. \label{eq:2.2.7}
\end{align}
Notice that when $\phi = 0$ ($\phi = \pi$), that is, $U = \sigma_{3}$ ($U = -\sigma_{3}$), the boundary conditions become type ($0,\pi$) (type ($\pi, 0$)) boundary conditions in the interval case and again lead to $j_{0}(0) = 0 = j_{0}(L)$.
Thus in the circle case these two ``points'' $U = \sigma_{3}$ and $-\sigma_{3}$, which correspond to the north pole $\phi = 0$ and the south pole $\phi = \pi$ of $S^{2}$, respectively, must be removed from the parameter space $S^{2}$.
The resulting parameter space is thus isomorphic to a non-compact two-dimensional cylinder.
In summary the boundary conditions compatible with $\mathscr{N}=2$ supersymmetry have a two-parameter family, which can be written as
\begin{subequations}
\begin{align}
	\begin{bmatrix}
	\phi_{0}(L) \\
	(Q_{0}\phi_{0})(L)
	\end{bmatrix}
&= 	{\rm e}^{i\theta}
	\begin{bmatrix}
	\tan(\phi/2) 	& 0 \\
	0 			& \cot(\phi/2)
	\end{bmatrix}
	\begin{bmatrix}
	\phi_{0}(0) \\
	(Q_{0}\phi_{0})(0)
	\end{bmatrix}, \label{eq:2.2.8a}\\
	\begin{bmatrix}
	\phi_{1}(L) \\
	(Q_{0}^{\dagger}\phi_{1})(L)
	\end{bmatrix}
&= 	{\rm e}^{i\theta}
	\begin{bmatrix}
	\cot(\phi/2) 	& 0 \\
	0 			& \tan(\phi/2)
	\end{bmatrix}
	\begin{bmatrix}
	\phi_{1}(0) \\
	(Q_{0}^{\dagger}\phi_{1})(0)
	\end{bmatrix}, \label{eq:2.2.8b}
\end{align}
\end{subequations}
where $0\leq\theta<2\pi$ and $0<\phi<\pi$.
In practical calculations it is convenient to introduce a real parameter $\eta$ defined as
\begin{align}
\mathrm{e}^{\eta}
:= 	\tan\left(\frac{\phi}{2}\right), \quad
-\infty<\eta<\infty.
\end{align}
Before closing this section, we should make a comment on physical meanings of these two parameters $\theta$ and $\eta$.
As is well-known, $\theta$ corresponds to the magnetic flux penetrating through the circle.
On the other hand, as shown in \cite{Griffiths:1993}, boundary conditions with nonzero $\eta$ corresponds to the presence of $\delta^{\prime}$-singularity at the junction point $x=0$.

%%%%% REFACTORIZATION OF HAMILTONIANS %%%%%
\section{Refactorization of Hamiltonians} \label{sec:refactorization}
As already mentioned in Section \ref{sec:intro}, quantum mechanical supersymmetry plays an essential role to generate massive Kaluza-Klein particles in higher-dimensional field theory.
It has been shown that in 5d gravity \textit{two} $\mathscr{N}=2$ quantum mechanical SUSYs are needed in order 
for the ``Higgs'' mechanism to generate massive spin-2 particles \cite{Lim:2007}.
A crucial ingredient of this coexistence of two quantum mechanical SUSYs is the refactorization of Hamiltonians.
Thus it would be natural to guess that in a higher-dimensional spin-$N$ field theory there would exist a hierarchy of $N$ SUSYs in the 4d mass spectrum, whose typical structure must be as follows:
\begin{center}
\begin{tabular}{cccc}
$H_{0}$ 	&$= 	Q_{0}^{\dagger}Q_{0}$ 	& 								& \\
$H_{1}$ 	&$= 	Q_{0}Q_{0}^{\dagger}$ 	&$= 	Q_{1}^{\dagger}Q_{1} + c_{1}$ 	& \\
$H_{2}$ 	& 						&$= 	Q_{1}Q_{1}^{\dagger} + c_{1}$ 	&$= 	Q_{2}^{\dagger}Q_{2} + c_{1} + c_{2}$ \\
$H_{3}$ 	& 						& 								&$= 	Q_{2}Q_{2}^{\dagger} + c_{1} + c_{2}$ \\
$\vdots$	& 						& 								&$\vdots$
\end{tabular}
\end{center}
where the $n$-th supercharge and its adjoint are assumed to be of the form
\begin{subequations}
\begin{align}
Q_{n}
&= 	{\rm e}^{-W_{n}(x)}\frac{{\rm d}}{{\rm d}x}{\rm e}^{+W_{n}(x)}
= 	\frac{\rm d}{{\rm d}x} + W_{n}^{'}(x), \label{eq:3.1a}\\
Q_{n}^{\dagger}
&= 	-{\rm e}^{+W_{n}(x)}\frac{{\rm d}}{{\rm d}x}{\rm e}^{-W_{n}(x)}
= 	- \frac{\rm d}{{\rm d}x} + W_{n}^{'}(x), \label{eq:3.1b}
\end{align}
\end{subequations}
and $c_{n}$ is a real constant.
In the context of higher-dimensional field theory, $W_{n}$ and $c_{n}$ would correspond to the warp factor and the cosmological constant on 3-branes, respectively.

In this section we solve the refactorization condition of Hamiltonians 
in the case of $c_{n}=0$ and construct a hierarchy of supersymmetry.

%%%%% REFACTORIZATION OF HAMILTONIANS %%%%%
\subsection{Refactorization of Hamiltonians}
Although in this paper we will focus on the case that all the constant shifts $c_{n}$ are zero, it may be instructive to keep $c_{n}$ to be nonzero in order to distinguish our refactorization method and the conventional one, which is used to solve the Schr\"odinger equation by the method of shape invariance.

The refactorization condition for the $n$-th Hamiltonian $Q_{n-1}Q_{n-1}^{\dagger} = Q_{n}^{\dagger}Q_{n} + c_{n}$ can be written into the following form
\begin{align}
(W_{n-1}^{\prime})^{2} + W_{n-1}^{\prime\prime}
= 	(W_{n}^{\prime})^{2} - W_{n}^{\prime\prime} + c_{n}. \label{eq:3.2}
\end{align}
This is a recursion relation known as the ladder equation in the context of parasupersymmetric or higher-derivative supersymmetric quantum mechanics \cite{Rubakov:1988,Andrianov1:1991,Andrianov2:1991,Andrianov:1993,Andrianov:1994,Andrianov:1995,FernandezC:1996}.
Our task is to solve the equation \eqref{eq:3.2} with respect to $W_{n}$ and to recursively define the $n$-th superpotential.
The nonlinear differential equation \eqref{eq:3.2} is the Riccati equation in terms of $W_{n}$ so that it can be linearized as follows:
\begin{align}
Q_{n-1}Q_{n-1}^{\dagger}\mathrm{e}^{-W_{n}} = c_{n}\mathrm{e}^{-W_{n}}, \label{eq:3.3}
\end{align}
or equivalently
\begin{align}
H_{n}\mathrm{e}^{-W_{n}} = \left(\sum_{i=1}^{n}c_{i}\right)\mathrm{e}^{-W_{n}}. \label{eq:3.4}
\end{align}
This is nothing but the Schr\"odinger equation for the $n$-th Hamiltonian.
Noting that the spectrum of $n$-th Hamiltonian is bounded from below by the constant $\sum_{i=1}^{n}c_{i}$, we see that Eq.\eqref{eq:3.4} is the Schr\"odinger equation for the ground state.

When $c_{n}=0$ it is easy to solve the equation \eqref{eq:3.3} with the result
\begin{align}
W_{n}
= 	- W_{n-1}
	- \ln\left\{\alpha_{n-1} + \beta_{n-1}\int_{x_{0}}^{x}\!\!\!
	{\rm d}y~{\rm e}^{-2W_{n-1}(y)}\right\}, \label{eq:3.5}
\end{align}
where $\alpha_{n}$ and $\beta_{n}$ are integration constants.
$x_{0}$ is an arbitrary point placed on the interval $(0,L)$.
Since in this paper we concentrate on finite superpotentials even at the boundaries, it is convenient to choose $x_{0}$ as $x_{0} = 0$ and $\beta_{n-1}$ as $\beta_{n-1} = \left[\int_{0}^{L}\!\!\mathrm{d}y\exp(-2W_{n-1})\right]^{-1}$.
We note that a constant shift of the superpotentials has no effect on the Hamiltonians.
With these choices, the parameter $\alpha_{n-1}$ is limited to the ranges $\alpha_{n-1}<-1$ and $0<\alpha_{n-1}$ for the well-definedness of $W_{n}^{\prime}$.
Thus, once given a quantum mechanical system, we can always 
construct an infinite hierarchy of Hamiltonians.

Notice that the result \eqref{eq:3.5} coincides with the so-called isospectral deformations of the Hamiltonian \cite{Abraham:1980,Baye:1987,Amado:1988}.

%%%%% THREE-TERM RECURRENCE RELATION %%%%%
\subsection{Three-term recurrence relation for nonzero-modes}
Let $\phi_{n}^{(l)}$ be the energy eigenfunction of $l$-th excited states for the $n$-th Hamiltonian.
Then, we have the three-term recurrence relation for quantum mechanical systems with $N$ SUSYs:
\begin{align}
\phi_{n+2}^{(l)}
= 	-\phi_{n}^{(l)}
	+ \frac{1}{\sqrt{E_{l}}}(W_{n}^{'} + W_{n+1}^{'})\phi_{n+1}^{(l)}, 
	\label{eq:3.8}
\end{align}
which follows from the SUSY relations $\sqrt{E_{l}}\phi_{n+2}^{(l)}
=Q_{n+1}\phi_{n+1}^{(l)},\ \sqrt{E_{l}}\phi_{n}^{(l)}
=Q_{n}^{\dagger}\phi_{n+1}^{(l)}$ and the identity $Q_{n+1} 
= -Q_{n}^{\dagger} + W_{n}^{'} + W_{n+1}^{'}$.
Notice that when $\beta_{n+1} = 0$, $\phi_{n+2}^{(l)}$ just reduces to the (opposite sign of) energy eigenfunction $\phi_{n}^{(l)}$.

%%%%% ZERO-MODE %%%%%
\subsection{Zero-mode}
\begin{figure*}[t]
	\begin{center}
		\begin{tabular}{cc}
		\includegraphics[scale=.9]{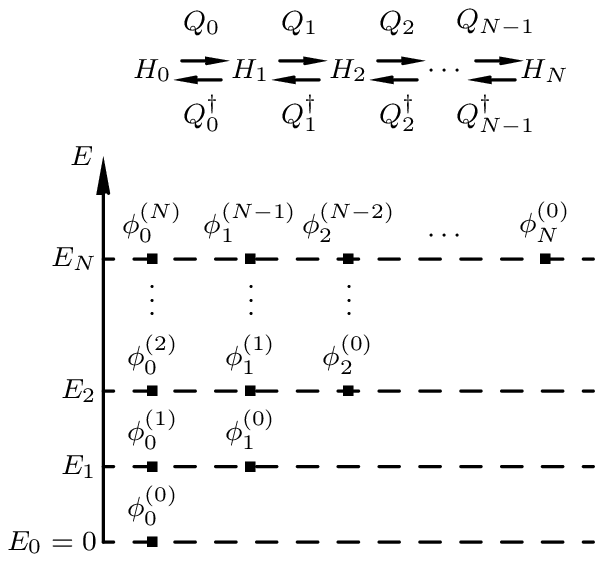}
		& \includegraphics[scale=.9]{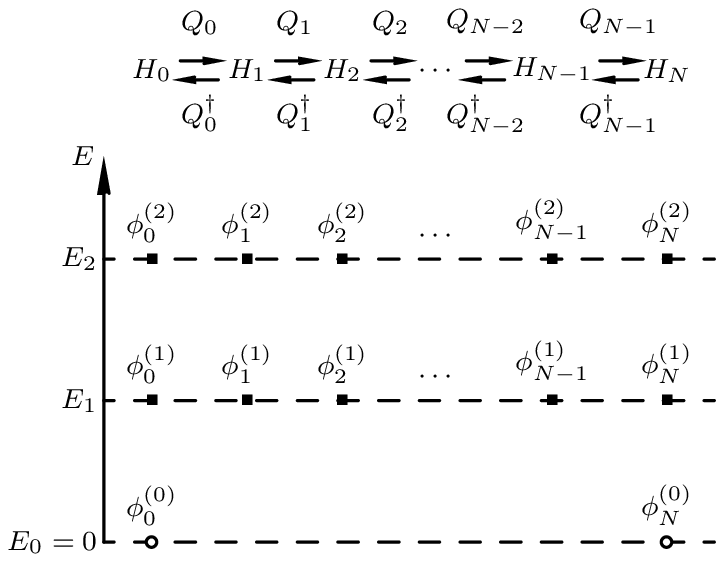}\\
		(a) 	& (b)
		\end{tabular}
		\caption{(a) Typical spectrum of quantum system constructed by the conventional refactorization method with $W_{n} = -\ln\phi_{n}^{(0)}$. 
		(b) Typical spectrum of quantum system with $N$ SUSYs.}	
		\label{fig:level_N}
	\end{center}
\end{figure*}

Next we will show that the zero-mode functions $\phi_{n}^{(0)}$ for $0<n<N$ cannot exist in general in a quantum mechanical system with $N$ SUSYs.
To this end, suppose that we have constructed a set of $N+1$ isospectral Hamiltonians using the refactorization method.
Since the $n$-th Hamiltonian $H_{n}$ can be written in two ways as $H_{n} = Q_{n-1}Q_{n-1}^{\dagger} = Q_{n}^{\dagger}Q_{n}$, $\phi_{n}^{(0)}(x)$ with $n=1,\cdots,N-1$ has to satisfy the equations
\begin{align}
Q_{n-1}^{\dagger}\phi_{n}^{(0)} = 0 = Q_{n}\phi_{n}^{(0)}, \label{eq:3.9}
\end{align}
or equivalently
\begin{align}
\left(
\frac{\mathrm{d}}{\mathrm{d}x} - W_{n-1}^{\prime}
\right)\phi_{n}^{(0)}
= 	0
= 	\left(
	\frac{\mathrm{d}}{\mathrm{d}x} - W_{n-1}^{\prime}
	- \frac{\beta_{n-1} {\rm e}^{-2W_{n-1}}}
	{\alpha_{n-1} + \beta_{n-1}
	\int_{x_{0}}^{x}\!\!\mathrm{d}y~\mathrm{e}^{-2W_{n-1}}}
	\right)
	\phi_{n}^{(0)}. \label{eq:3.10}
\end{align}
Obviously, there is no nontrivial solution to these two different equations except for the case $\beta_{n-1}=0$. When $\beta_{n-1} = 0$, the $(n+1)$-th Hamiltonian $H_{n+1} = Q_{n+1}^{\dagger}Q_{n+1}$ comes to be identical to the $(n-1)$-th Hamiltonian, which has no interest for us.
Therefore there is no nontrivial solution to \eqref{eq:3.9}.
We thus conclude that the zero-mode solutions consistent with $N$ SUSYs can exist {\em at most} only for the case $n=0$ and $N$.
The ground-state energy eigenfunction for $H_{N}$ is obtained by solving the equation $Q_{N-1}^{\dagger}\phi_{N}^{(0)} = 0$, which can be easily integrated with the result
\begin{align}
\phi_{N}^{(0)}(x)
= 	C{\rm e}^{+W_{N-1}(x)}, \label{eq:3.11}
\end{align}
where $C$ is the normalization constant.
If $\phi_{N}^{(0)}$ turns out to be non-normalizable or not to obey the boundary conditions, only a single zero-mode $\phi_{0}^{(0)}$ exists.
Typical spectrum of a quantum mechanical system with $N$ SUSYs is shown in Figure \ref{fig:level_N}.

%%%%% BOUNDARY CONDITIONS %%%%%
\section{Hierarchy of QM SUSYs} \label{sec:hierarchy}
In the previous section, we have not discussed boundary conditions
compatible with $N$ SUSYs.
In this section we will investigate whether it is possible to construct a hierarchical SUSY without conflicting with the hermiticity of each Hamiltonian.
In the subsequent subsections we will study this hierarchical SUSY on an interval and on a circle separately.

%%%%% HIERARCHY ON AN INTERVAL %%%%%
\subsection{Hierarchy on an interval} \label{subsec:hierarchy_interval}
Let us first study a hierarchical SUSY on an interval.
As a first step let us consider the boundary conditions consistent with 2 SUSYs.
Inserting the supersymmetric relations $Q_{1}\phi_{1} = \sqrt{E}\phi_{2}$ and $Q_{1}^{\dagger}\phi_{2} = \sqrt{E}\phi_{1}$ into the equation \eqref{eq:2.1.3a} we have
\begin{subequations}
\begin{alignat}{2}
\label{eq:28a}
\phi_{0}:~&
0
&=& 	\sin\left(\frac{\theta_{i}}{2}\right)\phi_{0}(x_{i})
	+ L_{0}\cos\left(\frac{\theta_{i}}{2}\right)(Q_{0}\phi_{0})(x_{i}), \\
\label{eq:28b}
\phi_{1}:~&
0
&=& 	\sin\left(\frac{\theta_{i}}{2}\right)(Q_{0}^{\dagger}\phi_{1})(x_{i})
+ EL_{0}\cos\left(\frac{\theta_{i}}{2}\right)\phi_{1}(x_{i}), \\
\label{eq:28c}
\phi_{2}:~&
0
&=& 	\sin\left(\frac{\theta_{i}}{2}\right)(W_{0}^{\prime}+W_{1}^{\prime})(x_{i})(Q_{1}^{\dagger}\phi_{2})(x_{i}) \nonumber\\
&&& 	+ E\left\{
	-\sin\left(\frac{\theta_{i}}{2}\right)\phi_{2}(x_{i})
	+ L_{0}\cos\left(\frac{\theta_{i}}{2}\right)(Q_{1}^{\dagger}\phi_{2})(x_{i})
	\right\},
\end{alignat}
\end{subequations}
where the third equation follows from Eq.\eqref{eq:28b} with the identity $Q_{0}^{\dagger} = -Q_{1}+W_{0}^{\prime}+W_{1}^{\prime}$.
Now it is obvious that there is no possible boundary conditions independent of $E$ except for the choice $\theta_{i} = 0$.
Thus the boundary conditions consistent with 2 SUSYs are uniquely determined as follows:
\begin{subequations}
\begin{align}
\label{eq:29a}
(Q_{0}\phi_{0})(x_{i})
&= 0, \\
\label{eq:29b}
\phi_{1}(x_{i})
&= 0, \\
\label{eq:29c}
(Q_{1}^{\dagger}\phi_{2})(x_{i})
&= 0.
\end{align}
\end{subequations}
It is easy to show that there is no possible boundary conditions consistent with a hierarchy of $N$ SUSYs for $N\geq3$.
Thus, we conclude that, at most, three successive quantum mechanical systems on an interval can be supersymmetric in a hierarchy of QM SUSYs.

%%%%% HIERARCHY ON A CIRCLE %%%%%
\subsection{Hierarchy on a circle} \label{subsec:hierarchy_circle}
Let us next study a hierarchical SUSY on a circle.
As mentioned before in this paper we focus on finite superpotentials on the whole domain.
When $W_{0}$ is finite, the finite $(n+1)$-th superpotential $W_{n+1}$ is recursively defined as
\begin{align}
W_{n+1}(x)
&= 	-W_{n}(x)
	-\ln\left[\alpha_{n} + \beta_{n}\int_{0}^{x}\!\!\!\mathrm{d}y~\mathrm{e}^{-2W_{n}(y)}\right],
	\quad\text{for}\quad
	n=0,1,2,\cdots, \label{eq:4.2.1}
\end{align}
with
\begin{align}
\alpha_{n}<-1\ \ \textrm{or}\ \ 0<\alpha_{n}, \quad
\beta_{n}
= 	\left[\int_{0}^{L}\!\!\!\mathrm{d}x~\mathrm{e}^{-2W_{n}(x)}\right]^{-1}. \label{eq:4.2.2}
\end{align}
Since the hierarchy of $N$ SUSYs is just the assembly of $\mathscr{N}=2$ SUSYs, the boundary conditions in $H_{n}$--$H_{n+1}$ sector have to be of the form
\begin{subequations}
\begin{align}
	\begin{bmatrix}
	\phi_{n}(L) \\
	(Q_{n}\phi_{n})(L)
	\end{bmatrix}
&= 	{\rm e}^{i\theta_{n}}
	\begin{bmatrix}
	\mathrm{e}^{\eta_{n}} 	& 0 \\
	0 					& \mathrm{e}^{-\eta_{n}}
	\end{bmatrix}
	\begin{bmatrix}
	\phi_{n}(0) \\
	(Q_{n}\phi_{n})(0)
	\end{bmatrix}, \label{eq:4.2.3a}\\
	\begin{bmatrix}
	\phi_{n+1}(L) \\
	(Q_{n}^{\dagger}\phi_{n+1})(L)
	\end{bmatrix}
&= 	{\rm e}^{i\theta_{n}}
	\begin{bmatrix}
	\mathrm{e}^{-\eta_{n}} 	& 0 \\
	0 					& \mathrm{e}^{\eta_{n}}
	\end{bmatrix}
	\begin{bmatrix}
	\phi_{n+1}(0) \\
	(Q_{n}^{\dagger}\phi_{n+1})(0)
	\end{bmatrix}, \label{eq:4.2.3b}
\end{align}
\end{subequations}
with
\begin{align}
0\leq\theta_{n}<2\pi \quad\text{and}\quad -\infty<\eta_{n}<\infty. \label{eq:4.2.4}
\end{align}
For the sake of concreteness of the discussion, let us first consider 2 SUSYs in $H_{0}$--$H_{1}$--$H_{2}$ sector.
The point is whether there exists a well-defined parameter region to be consistent with two different boundary conditions for the wavefunction $\phi_{1}(x)$ of the middle Hamiltonian system $H_{1}$:
\begin{subequations}
\begin{align}
\phi_{1}(L)
&= 	\mathrm{e}^{i\theta_{0}-\eta_{0}}\phi_{1}(0), \label{eq:4.2.5a}\\
(Q_{0}^{\dagger}\phi_{1})(L)
&= 	\mathrm{e}^{i\theta_{0}+\eta_{0}}(Q_{0}^{\dagger}\phi_{1})(0), \label{eq:4.2.5b}
\end{align}
\end{subequations}
which come from Eq.\eqref{eq:4.2.3b} for $n=0$, and
\begin{subequations}
\begin{align}
\phi_{1}(L)
&= 	\mathrm{e}^{i\theta_{1}+\eta_{1}}\phi_{1}(0), \label{eq:4.2.6a}\\
(Q_{1}\phi_{1})(L)
&= 	\mathrm{e}^{i\theta_{1}-\eta_{1}}(Q_{1}\phi_{1})(0), \label{eq:4.2.6b}
\end{align}
\end{subequations}
which come from Eq.\eqref{eq:4.2.3a} for $n=1$.

\begin{figure}[t]
\begin{center}
\includegraphics{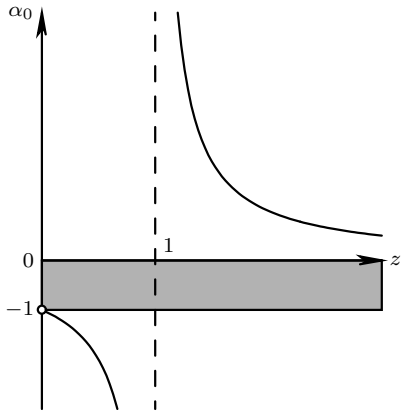}
\caption{Allowed region of the isospectral parameter $\alpha_{0}$ as a function of $z = \exp\left[-2\left(\eta_{0} + \int_{0}^{L}\!\mathrm{d}x~W_{0}^{\prime}(x)\right)\right]$, whose range is $0<z<\infty$.}
\label{fig:2}
\end{center}
\end{figure}

First, it is obvious that the parameters $\theta_{1}$ and $\eta_{1}$ have to be equal to $\theta_{0}$ and $-\eta_{0}$, respectively:
\begin{align}
\theta_{1} = \theta_{0}, \quad
\eta_{1} = -\eta_{0}. \label{eq:4.2.7}
\end{align}
Next, by adding Eqs.\eqref{eq:4.2.5a} and \eqref{eq:4.2.6b}
\begin{align}
\left(W_{0}^{\prime}(L) + W_{1}^{\prime}(L)\right)\phi_{1}(L)
&= 	\mathrm{e}^{i\theta_{0} + \eta_{0}}
	\left(W_{0}^{\prime}(0) + W_{1}^{\prime}(0)\right)\phi_{1}(0), \label{eq:4.2.8}
\end{align}
from which we find
\begin{align}
\mathrm{e}^{2\eta_{0}}
&= 	\frac{W_{0}^{\prime}(L) + W_{1}^{\prime}(L)}{W_{0}^{\prime}(0) + W_{1}^{\prime}(0)} \nonumber\\
&= 	\frac{\alpha_{0}}{1 + \alpha_{0}}
	\exp\left(-2\int_{0}^{L}\!\!\!\mathrm{d}x~W_{0}^{\prime}(x)\right), \label{eq:4.2.9}
\end{align}
where the last equality follows from Eq.\eqref{eq:4.2.1}.
Thus in order to implement the two boundary conditions the isospectral parameter $\alpha_{0}$ has to be tuned as
\begin{align}
{\alpha_{0}}^{-1}
&= 	\exp\left[-2\left(\eta_{0} + \int_{0}^{L}\!\!\!\mathrm{d}x~W_{0}^{\prime}(x)\right)\right]  - 1. \label{eq:4.2.10}
\end{align}
Notice that once the parameters $\eta_{1}$ and $\alpha_{0}$ are tuned as Eq.\eqref{eq:4.2.7} and Eq.\eqref{eq:4.2.10}, the following identity holds
\begin{align}
\eta_{1} + \int_{0}^{L}\!\!\!\mathrm{d}x~W_{1}^{\prime}(x)
&= 	\eta_{0} + \int_{0}^{L}\!\!\!\mathrm{d}x~W_{0}^{\prime}(x). \label{eq:4.2.11}
\end{align}
The above procedure can be easily continued to arbitrary $n$.
The resulting boundary conditions are as follows:
\begin{subequations}
\begin{align}
\phi_{n}(L)
&= 	\mathrm{e}^{i\theta_{0} \pm \eta_{0}}\phi_{n}(0), \label{eq:4.2.12a}\\
(Q_{n}\phi_{n})(L)
&= 	\mathrm{e}^{i\theta_{0} \mp \eta_{0}}(Q_{n}\phi_{n})(0), \label{eq:4.2.12b}
\end{align}
\end{subequations}
where $+$ ($-$) sign for $n=0,2,4\cdots$ ($n=1,3,5\cdots$).
The isospectral parameters are tuned as
\begin{align}
{\alpha_{n}}^{-1}
= 	\exp\left[-2\left(\eta_{0} + \int_{0}^{L}\!\!\!\mathrm{d}x~W_{0}^{\prime}(x)\right)\right]  - 1,
\quad
n=0,1,2,\cdots, \label{eq:4.2.13}
\end{align}
where $\alpha_{n}$ takes a desired value of $\alpha_{n}<-1$ or $\alpha_{n}>0$ (see Fig.\ref{fig:2}), as it should be.
We thus conclude that starting from any quantum mechanical system on a circle we can systematically construct an infinite hierarchy of QM SUSYs.
We should emphasize the difference between the hierarchy on an interval and on a circle.
In the hierarchy on an interval, at most, three successive quantum mechanical systems can be supersymmetric with the
unique boundary conditions \eqref{eq:29a}$-$\eqref{eq:29c}.
On the other hand, in the hierarchy on a circle, we can obtain an infinite tower of quantum mechanical systems whose successive two systems form an $\mathscr{N} =2$ SUSY with the boundary conditions \eqref{eq:4.2.12a} and \eqref{eq:4.2.12b}, which are specified by two parameters $\theta_{0}, \eta_{0}$.

%%%%% CONCLUSION %%%%%
\section{Conclusions and discussions} \label{sec:conclusion}
In this paper we have clarified the possible boundary conditions in $\mathscr{N}=2$ supersymmetric quantum mechanics on a finite domain $(0,L)$ without conflicting with the conservation of probability current.
Allowed boundary conditions in $\mathscr{N}=2$ supersymmetric quantum mechanics are limited to the so-called scale-independent subfamily of the $U(2)$ family of boundary conditions.
We also studied the hierarchy of $N$ SUSYs and showed that in an interval case it is not possible to construct beyond 2 SUSYs.
On the other hand, in a circle case it is possible to construct an infinite hierarchy of supersymmetries by tuning the isospectral parameters $\alpha_{n}$.

Let us close with some remarks.
\begin{enumerate}
\item \textbf{\mathversion{bold}Loop effects of $\eta$}.
	We show that in $\mathscr{N}=2$ supersymmetric quantum mechanics on a circle it is possible to introduce two parameter $\theta$ and $\eta$ into the boundary conditions.
	As mentioned in Section \ref{sec:BC}, $\theta$ corresponds to the magnetic flux penetrating through the circle and nonzero $\eta$ corresponds to the presence of the $\delta^{\prime}$-singularity at the junction point $x=0$.
	In higher-dimensional gauge theory compactified on a circle it is widely known that the twisted boundary conditions give rise to gauge symmetry/supersymmetry breaking known as the Hosotani/Schark-Schwarz mechanism.
	However, the effect of the presence of $\eta$ is not fully understood yet.
	It is interesting to investigate the loop effects of the parameter $\eta$ in five-dimensional gauge theory with a single extra dimension compactified on a circle.
	We will address this issue elsewhere.

\item \textbf{Integrable models}.
	As opposed to the shape invariant method, the techniques developed in this paper cannot be used to solve the Schr\"odinger equation.
	However, once given a solvable model, it is possible to generate an infinite tower of isospectral solvable models with nontrivial potential energy terms.

\item \textbf{\mathversion{bold}Spin-$N$ field theory}.
	In this paper we formulate a systematic description for constructing the hierarchy of $N$ SUSYs and show that in an interval case it is not possible to construct beyond 2 SUSYs.
	Since it seems a necessary condition in order to generate massive Kaluza-Klein particles, one might expect that it is possible to prove some kind of no-go theorem of the ``Higgs'' mechanism for spin-$N$ ($\geq3$) particle in the context of five-dimensional field theory with a single extra dimension compactified on an interval.
	However this is an open question.

\item \textbf{\mathversion{bold}Relax to $\mathcal{PT}$-symmetry}.
	Recently, a considerable number of studies have been made on non-hermitian $\mathcal{PT}$-symmetric quantum mechanics (see for recent review \cite{Bender:2007}).
	It is known that the conventional hermiticity condition on Hamiltonian is the sufficient condition for the real and lower bounded spectra and can be replaced by the weaker condition of the $\mathcal{PT}$-symmetry of Hamiltonian.
	In this paper we impose the hermiticity of Hamiltonian, however, it is interesting to relax the hermiticity condition to $\mathcal{PT}$-symmetric one.
	But it is not clear to the authors how to treat the $\mathcal{PT}$-symmetry into the boundary conditions.
\end{enumerate}

%%%%% ACKNOWLEDGMENT %%%%%
\section*{Acknowledgment}
One of the authors (SO) would like to thank S. Odake for very useful discussions and his hospitality.
This work is supported in part by the Grant-in-Aid for Scientific 
Research (No.18540275) 
by the Japanese Ministry of Education, Science, Sports and Culture.

%%%%% REFERENCES %%%%%
\addcontentsline{toc}{section}{\textsf{References}}


\begin{thebibliography}{3}
\expandafter\ifx\csname natexlab\endcsname\relax\def\natexlab#1{#1}\fi
\expandafter\ifx\csname bibnamefont\endcsname\relax
  \def\bibnamefont#1{#1}\fi
\expandafter\ifx\csname bibfnamefont\endcsname\relax
  \def\bibfnamefont#1{#1}\fi
\expandafter\ifx\csname citenamefont\endcsname\relax
  \def\citenamefont#1{#1}\fi
\expandafter\ifx\csname url\endcsname\relax
  \def\url#1{\texttt{#1}}\fi
\expandafter\ifx\csname urlprefix\endcsname\relax\def\urlprefix{URL }\fi
\providecommand{\bibinfo}[2]{#2}
\providecommand{\eprint}[2][]{\url{#2}}
%1
\bibitem[1]{Witten:1981}
	\bibinfo{author}{\bibfnamefont{E.}~\bibnamefont{Witten}},
	\textit{\bibinfo{title}{Dynamical breaking of supersymmetry}},
	\href{http://dx.doi.org/10.1016/0550-3213(81)90006-7}
	{\bibinfo{journal}{Nucl. Phys.} \textbf{\bibinfo{volume}{B188}}
	(\bibinfo{year}{1981}) \bibinfo{pages}{513}}.
%2
\bibitem[2]{Cooper:1994}
	\bibinfo{author}{\bibfnamefont{F.}~\bibnamefont{Cooper}},
	\bibinfo{author}{\bibfnamefont{A.}~\bibnamefont{Khare}} \bibnamefont{and}
	\bibinfo{author}{\bibfnamefont{U.}~\bibnamefont{Sukhatme}},
	\textit{\bibinfo{title}{Supersymmetry and quantum mechanics}},
	\href{http://www.yukawa.kyoto-u.ac.jp/cgi-bin/spiface/find/hep/www?eprint=hep-th/9405029}
	{\bibinfo{journal}{Phys. Rep.} \textbf{\bibinfo{volume}{251}}
	(\bibinfo{year}{1995}) \bibinfo{pages}{267}},
	\href{http://arxiv.org/abs/hep-th/9405029}
	{{\tt arXiv:hep-th/9405029}}.
%3
\bibitem[3]{Wang:1990}
	\bibinfo{author}{\bibfnamefont{Q.}~\bibnamefont{Wang}},
	\bibinfo{author}{\bibfnamefont{U.~P.}~\bibnamefont{Sukhatme}},
	\bibinfo{author}{\bibfnamefont{W.-Y.}~\bibnamefont{Keung}} \bibnamefont{and}
	\bibinfo{author}{\bibfnamefont{T.~D.}~\bibnamefont{Imbo}},
	\textit{\bibinfo{title}{Solitons from supersymmetry}},
	\href{http://dx.doi.org/10.1142/S0217732390000603}
	{\bibinfo{journal}{Mod. Phys. Lett.} \textbf{\bibinfo{volume}{A5}}
	(\bibinfo{year}{1990}) \bibinfo{pages}{525}}.
%4
\bibitem[4]{Grant:1993}
	\bibinfo{author}{\bibfnamefont{A.~K.}~\bibnamefont{Grant}} \bibnamefont{and}
	\bibinfo{author}{\bibfnamefont{J.~L.}~\bibnamefont{Rosner}},
	\textit{\bibinfo{title}{Supersymmetric quantum mechanics and the Korteweg-de Vries hierarchy}},
	\href{http://dx.doi.org/10.1063/1.530543}
	{\bibinfo{journal}{J. Math. Phys.} \textbf{\bibinfo{volume}{35}}
	(\bibinfo{year}{1994}) \bibinfo{pages}{2142}},
	\href{http://arxiv.org/abs/hep-th/9304139}
	{{\tt arXiv:hep-th/9304139}}.
%5
\bibitem[5]{Rodrigues:1998}
	\bibinfo{author}{\bibfnamefont{R.}~\bibnamefont{de~Lima~Rodrigues}},
	\bibinfo{author}{\bibfnamefont{P.~B.}~\bibnamefont{de~Silva~Filho}} \bibnamefont{and}
	\bibinfo{author}{\bibfnamefont{A.~N.}~\bibnamefont{Vaidya}},
	\textit{\bibinfo{title}{SUSY QM and solitons from two coupled scalar fields in two-dimensions}},
	\href{http://dx.doi.org/10.1103/PhysRevD.58.125023}
	{\bibinfo{journal}{Phys. Rev.} \textbf{\bibinfo{volume}{D58}}
	(\bibinfo{year}{1998}) \bibinfo{pages}{125023}}.
%6
\bibitem[6]{GomesLima:2002}
	\bibinfo{author}{\bibfnamefont{V.}~\bibnamefont{Gomes~Lima}},
	\bibinfo{author}{\bibfnamefont{V.}~\bibnamefont{Silva~Santos}} \bibnamefont{and}
	\bibinfo{author}{\bibfnamefont{R.}~\bibnamefont{De~Lima~Rodrigues}},
	\textit{\bibinfo{title}{On the scalar potential models from the isospectral potential class}},
	\href{http://dx.doi.org/10.1016/S0375-9601(02)00446-2}
	{\bibinfo{journal}{Phys. Lett.} \textbf{\bibinfo{volume}{A298}}
	(\bibinfo{year}{2002}) \bibinfo{pages}{91}},
	\href{http://arxiv.org/abs/hep-th/0204175}
	{{\tt arXiv:hep-th/0204175}}.
%7
\bibitem[7]{Dias:2002}
	\bibinfo{author}{\bibfnamefont{G.~S.}~\bibnamefont{Dias}},
	\bibinfo{author}{\bibfnamefont{E.~L.}~\bibnamefont{Graca}} \bibnamefont{and}
	\bibinfo{author}{\bibfnamefont{R.}~\bibnamefont{de~Lima~Rodrigues}},
	\textit{\bibinfo{title}{Stability equation and two component eigenmode for domain walls in a scalar potential model}},
	\href{http://dx.doi.org/10.1142/S0217751X07034994}
	{\bibinfo{journal}{Int. J. Mod. Phys.} \textbf{\bibinfo{volume}{A22}}
	(\bibinfo{year}{2007}) \bibinfo{pages}{731}},
	\href{http://arxiv.org/abs/hep-th/0205195}
	{{\tt arXiv:hep-th/0205195}}.
%8
\bibitem[8]{deLima:2003}
	\bibinfo{author}{\bibfnamefont{A.~F.}~\bibnamefont{de~Lima}} \bibnamefont{and}
	\bibinfo{author}{\bibfnamefont{R.}~\bibnamefont{de~Lima~Rodrigues}},
	\textit{\bibinfo{title}{q deformed kink solutions}},
	\href{http://dx.doi.org/10.1142/S0217751X06031375}
	{\bibinfo{journal}{Int. J. Mod. Phys.} \textbf{\bibinfo{volume}{A21}}
	(\bibinfo{year}{2006}) \bibinfo{pages}{3605}},
	\href{http://arxiv.org/abs/hep-th/0301114}
	{{\tt arXiv:hep-th/0301114}}.
%9
\bibitem[9]{Sukumar:1985}
	\bibinfo{author}{\bibfnamefont{C.~V.}~\bibnamefont{Sukumar}},
	\textit{\bibinfo{title}{Supersymmetric Quantum mechanics and the inverse scattering method}},
	\href{http://dx.doi.org/10.1088/0305-4470/18/15/021}
	{\bibinfo{journal}{J. Phys.} \textbf{\bibinfo{volume}{A18}}
	(\bibinfo{year}{1985}) \bibinfo{pages}{2937}}.
%10
\bibitem[10]{Sparenberg:1997}
	\bibinfo{author}{\bibfnamefont{J.-M.}~\bibnamefont{Sparenberg}} \bibnamefont{and}
	\bibinfo{author}{\bibfnamefont{D.}~\bibnamefont{Baye}},
	\textit{\bibinfo{title}{Inverse scattering with singular potentials: A supersymmetric approach}},
	\href{http://dx.doi.org/10.1103/PhysRevC.55.2175}
	{\bibinfo{journal}{Phys. Rev.} \textbf{\bibinfo{volume}{C55}}
	(\bibinfo{year}{1997}) \bibinfo{pages}{2175}}.
%11
\bibitem[11]{Baye:2004}
	\bibinfo{author}{\bibfnamefont{D.}~\bibnamefont{Baye}} \bibnamefont{and}
	\bibinfo{author}{\bibfnamefont{J.-M.}~\bibnamefont{Sparenberg}},
	\textit{\bibinfo{title}{Inverse scattering with supersymmetric quantum mechanics}},
	\href{http://dx.doi.org/10.1088/0305-4470/37/43/014}
	{\bibinfo{journal}{J. Phys.} \textbf{\bibinfo{volume}{A37}}
	(\bibinfo{year}{2004}) \bibinfo{pages}{10223}}.
%12
\bibitem[12]{Feinberg:2003}
	\bibinfo{author}{\bibfnamefont{J.}~\bibnamefont{Feinberg}},
	\textit{\bibinfo{title}{All about the static fermion bags in the Gross-Neveu model}},
	\href{http://dx.doi.org/10.1016/j.aop.2003.08.004}
	{\bibinfo{journal}{Ann. Phys.} \textbf{\bibinfo{volume}{309}}
	(\bibinfo{year}{2004}) \bibinfo{pages}{166}},
	\href{http://arxiv.org/abs/hep-th/0305240}
	{{\tt arXiv:hep-th/0305240}}.
%13
\bibitem[13]{Seeger:1998}
	\bibinfo{author}{\bibfnamefont{M.}~\bibnamefont{Seeger}} \bibnamefont{and}
	\bibinfo{author}{\bibfnamefont{M.}~\bibnamefont{Thies}},
	\textit{\bibinfo{title}{(1+1)-dimensional QCD with static quarks as supersymmetric quantum mechanics}},
	\href{http://dx.doi.org/10.1103/PhysRevD.58.027701}
	{\bibinfo{journal}{Phys. Rev.} \textbf{\bibinfo{volume}{D58}}
	(\bibinfo{year}{1998}) \bibinfo{pages}{027701}},
	\href{http://arxiv.org/abs/hep-th/9802060}
	{{\tt arXiv:hep-th/9802060}}.
%14
\bibitem[14]{Bergner:2007}
	\bibinfo{author}{\bibfnamefont{G.}~\bibnamefont{Bergner}},
	\bibinfo{author}{\bibfnamefont{T.}~\bibnamefont{Kaestner}},
	\bibinfo{author}{\bibfnamefont{S.}~\bibnamefont{Uhlmann}} \bibnamefont{and}
	\bibinfo{author}{\bibfnamefont{A.}~\bibnamefont{Wipf}},
	\textit{\bibinfo{title}{Low-dimensional supersymmetric lattice models}},
	\href{http://dx.doi.org/10.1016/j.aop.2007.06.010}
	{\bibinfo{journal}{Ann. Phys.} \textbf{\bibinfo{volume}{323}}
	(\bibinfo{year}{2008}) \bibinfo{pages}{946}},
	\href{http://arxiv.org/abs/0705.2212}
	{{\tt arXiv:0705.2212[hep-lat]}}.
%15
\bibitem[15]{Calogero:1971}
	\bibinfo{author}{\bibfnamefont{F.}~\bibnamefont{Calogero}},
	\textit{\bibinfo{title}{Solution of a three-body problem in one-dimension}},
	\href{http://dx.doi.org/10.1063/1.1664820}
	{\bibinfo{journal}{J. Math. Phys.} \textbf{\bibinfo{volume}{10}}
	(\bibinfo{year}{1969}) \bibinfo{pages}{2191}}.
%16
\bibitem[16]{Sutherland:1972}
	\bibinfo{author}{\bibfnamefont{B.}~\bibnamefont{Sutherland}},
	\textit{\bibinfo{title}{Exact Results for a Quantum Many-Body Problem in One Dimension. I\hspace{-.1em}I}},
	\href{http://dx.doi.org/10.1103/PhysRevA.5.1372 }
	{\bibinfo{journal}{Phys. Rev.} \textbf{\bibinfo{volume}{A5}}
	(\bibinfo{year}{1972}) \bibinfo{pages}{1372}}.
%17
\bibitem[17]{Moser:1975}
	\bibinfo{author}{\bibfnamefont{J.}~\bibnamefont{Moser}},
	\textit{\bibinfo{title}{Three integrable Hamiltonian systems connected with isospectral deformations}},
	\href{http://dx.doi.org/10.1016/0001-8708(75)90151-6}
	{\bibinfo{journal}{Adv. Math.} \textbf{\bibinfo{volume}{16}}
	(\bibinfo{year}{1975}) \bibinfo{pages}{197}}.
%18
\bibitem[18]{Gibbons:1998}
	\bibinfo{author}{\bibfnamefont{G.}~\bibfnamefont{W.}~\bibnamefont{Gibbons}} \bibnamefont{and}
	\bibinfo{author}{\bibfnamefont{P.}~\bibfnamefont{K.}~\bibnamefont{Townsend}},
	\textit{\bibinfo{title}{Black hole and Calogero models}},
	\href{http://dx.doi.org/10.1016/S0370-2693(99)00266-X}
	{\bibinfo{journal}{Phys. Lett.} \textbf{\bibinfo{volume}{B454}}
	(\bibinfo{year}{1999}) \bibinfo{pages}{187}},
	\href{http://arxiv.org/abs/hep-th/9812034}
	{{\tt arXiv:hep-th/9812034}}.
%19
\bibitem[19]{Meljanac:2006}
	\bibinfo{author}{\bibfnamefont{S.}~\bibnamefont{Meljanac}},
	\bibinfo{author}{\bibfnamefont{A.}~\bibnamefont{Samsarov}},
	\bibinfo{author}{\bibfnamefont{B.}~\bibnamefont{Basu-Mallick}} \bibnamefont{and}
	\bibinfo{author}{\bibfnamefont{K.}~\bibfnamefont{S.}~\bibnamefont{Gupta}},
	\textit{\bibinfo{title}{Quantization and conformal properties of a generalized Calogero model}},
	\href{http://dx.doi.org/10.1140/epjc/s10052-006-0163-9}
	{\bibinfo{journal}{Eur. Phys. J.} \textbf{\bibinfo{volume}{C49}}
	(\bibinfo{year}{2007}) \bibinfo{pages}{875}},
	\href{http://arxiv.org/abs/hep-th/0609111}
	{{\tt arXiv:hep-th/0609111}}.
%20
\bibitem[20]{point_singularity1}
	\bibinfo{author}{\bibfnamefont{T.}~\bibnamefont{Uchino}}
    \bibnamefont{and}
	\bibinfo{author}{\bibfnamefont{I.}~\bibnamefont{Tsutsui}},
	\textit{\bibinfo{title}{Supersymmetric quantum mechanics with 
	a point singularity}},
	\href{http://dx.doi.org/10.1140/epjc/s10052-006-0163-9}
	{\bibinfo{journal}{Nucl. Phys.} \textbf{\bibinfo{volume}{B662}}
	(\bibinfo{year}{2003}) \bibinfo{pages}{447}},
	\href{http://arxiv.org/abs/quant-ph/0210084}
	{{\tt arXiv:quant-ph/0210084}};
	%
	\textit{\bibinfo{title}{Supersymmetric quantum mechanics under 
	point singularities}},
	\href{http://dx.doi.org/10.1140/epjc/s10052-006-0163-9}
	{\bibinfo{journal}{J. Phys.} \textbf{\bibinfo{volume}{A36}}
	(\bibinfo{year}{2003}) \bibinfo{pages}{6493}},
	\href{http://arxiv.org/abs/hep-th/0302089}
	{{\tt arXiv:hep-th/0302089}}.
%21
\bibitem[21]{point_singularity2}
	\bibinfo{author}{\bibfnamefont{T.}~\bibnamefont{Nagasawa}},
	\bibinfo{author}{\bibfnamefont{M.}~\bibnamefont{Sakamoto}}
    \bibnamefont{and}
	\bibinfo{author}{\bibfnamefont{K.}~\bibnamefont{Takenaga}},
	\textit{\bibinfo{title}{Supersymmetry in quantum mechanics 
	with point interactions}},
	\href{http://dx.doi.org/10.1140/epjc/s10052-006-0163-9}
	{\bibinfo{journal}{Phys. Lett.} \textbf{\bibinfo{volume}{B562}}
	(\bibinfo{year}{2003}) \bibinfo{pages}{358}},
	\href{http://arxiv.org/abs/hep-th/0212192}
	{{\tt arXiv:hep-th/0212192}};
	%
	\textit{\bibinfo{title}{Supersymmetry and discrete transformations 
	on $S^1$ with point singularities}},
	\href{http://dx.doi.org/10.1140/epjc/s10052-006-0163-9}
	{\bibinfo{journal}{Phys. Lett} \textbf{\bibinfo{volume}{B583}}
	(\bibinfo{year}{2004}) \bibinfo{pages}{357}},
	\href{http://arxiv.org/abs/hep-th/0311043}
	{{\tt arXiv:hep-th/0311043}};
	%
	\textit{\bibinfo{title}{Extended supersymmetry and its reduction 
	on a circle with point singularities}},
	\href{http://dx.doi.org/10.1140/epjc/s10052-006-0163-9}
	{\bibinfo{journal}{J. Phys.} \textbf{\bibinfo{volume}{A38}}
	(\bibinfo{year}{2005}) \bibinfo{pages}{8053}},
	\href{http://arxiv.org/abs/hep-th/0505132}
	{{\tt arXiv:hep-th/0505132}}.
%22
\bibitem[22]{Lim:2005}
	\bibinfo{author}{\bibfnamefont{C.~S.}~\bibnamefont{Lim}},
	\bibinfo{author}{\bibfnamefont{T.}~\bibnamefont{Nagasawa}},
	\bibinfo{author}{\bibfnamefont{M.}~\bibnamefont{Sakamoto}} \bibnamefont{and}
	\bibinfo{author}{\bibfnamefont{H.}~\bibnamefont{Sonoda}},
	\textit{\bibinfo{title}{Supersymmetry in gauge theories with extra dimensions}},
	\href{http://dx.doi.org/10.1103/PhysRevD.72.064006}
	{\bibinfo{journal}{Phys. Rev.} \textbf{\bibinfo{volume}{D72}}
	(\bibinfo{year}{2005}) \bibinfo{pages}{064006}},
	\href{http://arxiv.org/abs/hep-th/0502022}
	{{\tt arXiv:hep-th/0502022}}.
%23
\bibitem[23]{hierarchy}
	\bibinfo{author}{\bibfnamefont{T.}~\bibnamefont{Nagasawa}}
    \bibnamefont{and}
	\bibinfo{author}{\bibfnamefont{M.}~\bibnamefont{Sakamoto}},	
	\textit{\bibinfo{title}{Higgsless gauge symmetry breaking 
	with a large mass hierarchy}},
	\href{http://dx.doi.org/10.1140/epjc/s10052-006-0163-9}
	{\bibinfo{journal}{Prog. Theor. Phys.} \textbf{\bibinfo{volume}{112}}
	(\bibinfo{year}{2004}) \bibinfo{pages}{629}},
	\href{http://arxiv.org/abs/hep-ph/0406024}
	{{\tt arXiv:hep-ph/0406024}}.
%24
\bibitem[24]{Georgi:2000}
	\bibinfo{author}{\bibfnamefont{H.}~\bibnamefont{Georgi}},
	\bibinfo{author}{\bibfnamefont{A.~K.}~\bibnamefont{Grant}} \bibnamefont{and}
	\bibinfo{author}{\bibfnamefont{G.}~\bibnamefont{Hailu}},
	\textit{\bibinfo{title}{Brane couplings from bulk loops}},
	\href{http://dx.doi.org/10.1016/S0370-2693(01)00408-7}
	{\bibinfo{journal}{Phys. Lett.} \textbf{\bibinfo{volume}{B506}}
	(\bibinfo{year}{2001}) \bibinfo{pages}{207}},
	\href{http://arxiv.org/abs/hep-ph/0012379}
	{{\tt arXiv:hep-ph/0012379}}.
%25
\bibitem[25]{Goldberger:2001}
	\bibinfo{author}{\bibfnamefont{W.~D.}~\bibnamefont{Goldberger}} \bibnamefont{and}
	\bibinfo{author}{\bibfnamefont{M.~B.}~\bibnamefont{Wise}},
	\textit{\bibinfo{title}{Renormalization group flows for brane couplings}},
	\href{http://dx.doi.org/10.1103/PhysRevD.65.025011}
	{\bibinfo{journal}{Phys. Rev.} \textbf{\bibinfo{volume}{D65}}
	(\bibinfo{year}{2002}) \bibinfo{pages}{025011}},
	\href{http://arxiv.org/abs/hep-th/0104170}
	{{\tt arXiv:hep-th/0104170}}.
%26
\bibitem[26]{Milton:2001}
	\bibinfo{author}{\bibfnamefont{K.~A.}~\bibnamefont{Milton}},
	\bibinfo{author}{\bibfnamefont{S.~D.}~\bibnamefont{Odintsov}} \bibnamefont{and}
	\bibinfo{author}{\bibfnamefont{S.}~\bibnamefont{Zerbini}},
	\textit{\bibinfo{title}{Bulk versus brane running couplings}},
	\href{http://dx.doi.org/10.1103/PhysRevD.65.065012}
	{\bibinfo{journal}{Phys. Rev.} \textbf{\bibinfo{volume}{D65}}
	(\bibinfo{year}{2002}) \bibinfo{pages}{065012}},
	\href{http://arxiv.org/abs/hep-th/0110051}
	{{\tt arXiv:hep-th/0110051}}.
%27
\bibitem[27]{Lim:2007}
	\bibinfo{author}{\bibfnamefont{C.~S.}~\bibnamefont{Lim}},
	\bibinfo{author}{\bibfnamefont{T.}~\bibnamefont{Nagasawa}},
	\bibinfo{author}{\bibfnamefont{S.}~\bibnamefont{Ohya}},
	\bibinfo{author}{\bibfnamefont{K.}~\bibnamefont{Sakamoto}} \bibnamefont{and}
	\bibinfo{author}{\bibfnamefont{M.}~\bibnamefont{Sakamoto}},
	\textit{\bibinfo{title}{Supersymmetry in 5d gravity}},
	\href{http://dx.doi.org/10.1103/PhysRevD.77.045020}
	{\bibinfo{journal}{Phys. Rev.} \textbf{\bibinfo{volume}{D77}}
	(\bibinfo{year}{2008}) \bibinfo{pages}{045020}},
	\href{http://arxiv.org/abs/0710.0170}
	{{\tt arXiv:0710.0170[hep-th]}};
%	
	\textit{\bibinfo{title}{Gauge-Fixing and Residual Symmetries 
	in Gauge/Gravity Theories with Extra Dimensions}},
	\href{http://dx.doi.org/10.1103/PhysRevD.77.045020}
	{\bibinfo{journal}{Phys. Rev.} \textbf{\bibinfo{volume}{D77}}
	(\bibinfo{year}{2008}) \bibinfo{pages}{065009}},
	\href{http://arxiv.org/abs/0710.0170}
	{{\tt arXiv:0801.0845[hep-th]}}.
%28
\bibitem[28]{Cheon:2000}
	\bibinfo{author}{\bibfnamefont{T.}~\bibnamefont{Cheon}},
	\bibinfo{author}{\bibfnamefont{T.}~\bibnamefont{F\"ul\"op}} \bibnamefont{and}
	\bibinfo{author}{\bibfnamefont{I.}~\bibnamefont{Tsutsui}},
	\textit{\bibinfo{title}{Symmetry, Duality, and Anholonomy of Point Interaction in One Dimension}},
	\href{http://dx.doi.org/10.1006/aphy.2001.6193}
	{\bibinfo{journal}{Ann. Phys.} \textbf{\bibinfo{volume}{294}}
	(\bibinfo{year}{2001}) \bibinfo{pages}{1}},
	\href{http://arxiv.org/abs/quant-ph/0008123}
	{{\tt arXiv:quant-ph/0008123}}.
%29
\bibitem[29]{Abraham:1980}
	\bibinfo{author}{\bibfnamefont{P.~B.}~\bibnamefont{Abraham}} \bibnamefont{and}
	\bibinfo{author}{\bibfnamefont{H.~E.}~\bibnamefont{Moses}},
	\textit{\bibinfo{title}{Changes in potentials due to changes in the point spectrum: Anharmonic oscillators with exact solutions}},
	\href{http://dx.doi.org/10.1103/PhysRevA.22.1333}
	{\bibinfo{journal}{Phys. Rev.} \textbf{\bibinfo{volume}{A22}}
	(\bibinfo{year}{1980}) \bibinfo{pages}{1333}}.
%30
\bibitem[30]{Baye:1987}
	\bibinfo{author}{\bibfnamefont{D.}~\bibnamefont{Baye}},
	\textit{\bibinfo{title}{Supersymmetry between deep and shallow nucleus-nucleus potentials}},
	\href{http://dx.doi.org/10.1103/PhysRevLett.58.2738}
	{\bibinfo{journal}{Phys. Rev. Lett.} \textbf{\bibinfo{volume}{58}}
	(\bibinfo{year}{1987}) \bibinfo{pages}{2738}}.
%31
\bibitem[31]{Amado:1988}
	\bibinfo{author}{\bibfnamefont{R.~D.}~\bibnamefont{Amado}},
	\textit{\bibinfo{title}{Phase-equivalent supersymmetric quantum-mechanical partners of the Coulomb potential}},
	\href{http://dx.doi.org/10.1103/PhysRevA.37.2277}
	{\bibinfo{journal}{Phys. Rev.} \textbf{\bibinfo{volume}{A37}}
	(\bibinfo{year}{1988}) \bibinfo{pages}{2277}}.
%32
\bibitem[32]{Griffiths:1993}
	\bibinfo{author}{\bibfnamefont{D.~J.}~\bibnamefont{Griffiths}},
	\textit{\bibinfo{title}{Boundary conditions at the derivative of a delta function}},
	\href{http://dx.doi.org/10.1088/0305-4470/26/9/021}
	{\bibinfo{journal}{J. Phys. A} \textbf{\bibinfo{volume}{26}}
	(\bibinfo{year}{1993}) \bibinfo{pages}{2265}}.
%33
\bibitem[33]{Rubakov:1988}
	\bibinfo{author}{\bibfnamefont{V.~A.}~\bibnamefont{Rubakov}} \bibnamefont{and}
	\bibinfo{author}{\bibfnamefont{V.~P.}~\bibnamefont{Spiridonov}},
	\textit{\bibinfo{title}{Parasupersymmetric quantum mechanics}},
	\href{http://dx.doi.org/10.1142/S0217732388001616}
	{\bibinfo{journal}{Mod. Phys. Lett.} \textbf{\bibinfo{volume}{A3}}
	(\bibinfo{year}{1988}) \bibinfo{pages}{1337}}.
%34
\bibitem[34]{Andrianov1:1991}
	\bibinfo{author}{\bibfnamefont{A.~A.}~\bibnamefont{Andrianov}} \bibnamefont{and}
	\bibinfo{author}{\bibfnamefont{M.~V.}~\bibnamefont{Ioffe}},
	\textit{\bibinfo{title}{From supersymmetric quantum mechanics to a parasupersymmetric one}},
	\href{http://dx.doi.org/10.1016/0370-2693(91)90263-P}
	{\bibinfo{journal}{Phys. Lett.} \textbf{\bibinfo{volume}{B255}}
	(\bibinfo{year}{1991}) \bibinfo{pages}{543}}.
%35
\bibitem[35]{Andrianov2:1991}
	\bibinfo{author}{\bibfnamefont{A.~A.}~\bibnamefont{Andrianov}},
	\bibinfo{author}{\bibfnamefont{M.~V.}~\bibnamefont{Ioffe}},
	\bibinfo{author}{\bibfnamefont{V.~P.}~\bibnamefont{Spiridonov}} \bibnamefont{and}
	\bibinfo{author}{\bibfnamefont{L.}~\bibnamefont{Vinet}},
	\textit{\bibinfo{title}{Parasupersymmetry and truncated supersymmetry in quantum mechanics}},
	\href{http://dx.doi.org/10.1016/0370-2693(91)91834-I}
	{\bibinfo{journal}{Phys. Lett.} \textbf{\bibinfo{volume}{B272}}
	(\bibinfo{year}{1991}) \bibinfo{pages}{297}}.
%36
\bibitem[36]{Andrianov:1993}
	\bibinfo{author}{\bibfnamefont{A.~A.}~\bibnamefont{Andrianov}},
	\bibinfo{author}{\bibfnamefont{M.~V.}~\bibnamefont{Ioffe}} \bibnamefont{and}
	\bibinfo{author}{\bibfnamefont{V.~P.}~\bibnamefont{Spiridonov}},
	\textit{\bibinfo{title}{Higher-derivative supersymmetry and the Witten index}},
	\href{http://dx.doi.org/10.1016/0375-9601(93)90137-O}
	{\bibinfo{journal}{Phys. Lett.} \textbf{\bibinfo{volume}{A174}}
	(\bibinfo{year}{1993}) \bibinfo{pages}{273}},
	\href{http://arxiv.org/abs/hep-th/9303005}
	{{\tt arXiv:hep-th/9303005}}.
%37
\bibitem[37]{Andrianov:1994}
	\bibinfo{author}{\bibfnamefont{A.~A.}~\bibnamefont{Andrianov}},
	\bibinfo{author}{\bibfnamefont{F.}~\bibnamefont{Cannata}},
	\bibinfo{author}{\bibfnamefont{J.-P.}~\bibnamefont{Dedonder}} \bibnamefont{and}
	\bibinfo{author}{\bibfnamefont{M.~V.}~\bibnamefont{Ioffe}},
	\textit{\bibinfo{title}{Second order derivative supersymmetry, Q deformations and scattering problem}},
	\href{http://dx.doi.org/10.1142/S0217751X95001261}
	{\bibinfo{journal}{Int. J. Mod. Phys.} \textbf{\bibinfo{volume}{A10}}
	(\bibinfo{year}{1995}) \bibinfo{pages}{2683}},
	\href{http://arxiv.org/abs/hep-th/9404061}
	{{\tt arXiv:hep-th/9404061}}.
%38
\bibitem[38]{Andrianov:1995}
	\bibinfo{author}{\bibfnamefont{A.~A.}~\bibnamefont{Andrianov}},
	\bibinfo{author}{\bibfnamefont{M.~V.}~\bibnamefont{Ioffe}} \bibnamefont{and}
	\bibinfo{author}{\bibfnamefont{D.~N.}~\bibnamefont{Nishnianidze}},
	\textit{\bibinfo{title}{Polynomial supersymmetry and dynamical symmetries in quantum mechanics}},
	\href{http://dx.doi.org/10.1007/BF02068745}
	{\bibinfo{journal}{Theor. Math. Phys.} \textbf{\bibinfo{volume}{104}}
	(\bibinfo{year}{1995}) \bibinfo{pages}{1129}}.
%39
\bibitem[39]{FernandezC:1996}
	\bibinfo{author}{\bibfnamefont{D.~J.}~\bibnamefont{Fern\'andez C.}},
	\textit{\bibinfo{title}{SUSUSY quantum mechanics}},
	\href{http://dx.doi.org/10.1142/S0217751X97000232}
	{\bibinfo{journal}{Int. J. Mod. Phys.} \textbf{\bibinfo{volume}{A12}}
	(\bibinfo{year}{1997}) \bibinfo{pages}{171}},
	\href{http://arxiv.org/abs/quant-ph/9609009}
	{{\tt arXiv:quant-ph/9609009}}.
%40
\bibitem[40]{Bender:2007}
	\bibinfo{author}{\bibfnamefont{C.~M.}~\bibnamefont{Bender}},
	\textit{\bibinfo{title}{Making sense of non-Hermitian Hamiltonians}},
	\href{http://dx.doi.org/10.1088/0034-4885/70/6/R03}
	{\bibinfo{journal}{Rep. Prog. Phys.} \textbf{\bibinfo{volume}{70}}
	(\bibinfo{year}{2007}) \bibinfo{pages}{947}},
	\href{http://arxiv.org/abs/hep-th/0703096}
	{{\tt arXiv:hep-th/0703096}}.
\end{thebibliography}
\end{document}